\documentclass[12pt,english,aps,manuscript]{article}
\usepackage{setspace}
\usepackage{geometry}
\usepackage{graphicx}
\usepackage{amsmath}
\usepackage{amssymb}

\begin{document}

\doublespacing 

\newcommand{\Ka}{K\"ahler }
\newcommand{\beq}{\begin{equation}}
\newcommand{\eeq}{\end{equation}}
\newcommand{\ord}{\mathcal{O}}

\begin{center}
\textbf{\Large Physical Vacua in IIB Compactifications with a Single \Ka Modulus}
\par\end{center}{\Large \par}

\begin{center}
\vspace{0.1cm}
\par\end{center}

\begin{center}
{\large Senarath de Alwis$^{\dagger}$ and Kevin Givens$^{\ddag}$ }
\par\end{center}{\large \par}

\begin{center}
Physics Department, University of Colorado \\
Boulder, CO 80309 USA
\par\end{center}

\begin{center}
\vspace{0.05cm}
\par\end{center}

\begin{center}
\textbf{Abstract}
\par\end{center}

We search for phenomenologically viable vacua of IIB string flux compactifications on Calabi-Yau orientifolds with a single \Ka modulus.   We perform both analytic studies and numerical searches in order to find models with de Sitter vacua and TeV-scale SUSY particle phenomenology.    
\vfill

$^{\dagger}$ {\small e-mail: dealwiss@colorado.edu}{\small \par}
$^{\ddag}$ {\small e-mail: kevin.givens@colorado.edu}{\small \par}

\eject

\vspace{5 mm}

\tableofcontents
\section{Introduction}

The search for physically plausible four dimensional vacua represents a preeminent goal of contemporary research in string theory.  The challenges endemic to this search originate principally from the fact that string theory is a ten dimensional theory that must be compactified to four dimensions.  The process of compactification necessarily introduces moduli fields that, from the standpoint of 4D effective field theory, must be stabilized with acceptable masses and vacuum expectation values.  For the case of IIB string theory, the general procedure for addressing these questions by using internal fluxes and non-perturbative terms has recently been developed.  For reviews see \cite{Grana:2005jc} and \cite{Douglas:2006es}.  

One of the principal drawbacks of an early model, the KKLT scenario\cite{Kachru:2003aw}, is that the moduli are \textit{a priori} stabilized at values producing a negative cosmological constant and that supersymmetry (SUSY) remains unbroken.  In order to achieve a de Sitter minimum the authors introduce $\overline{D3}$ branes into the compactified volume. This uplifts the scalar potential to a positive value and breaks supersymmetry. However, from a four dimensional supergravity (SUGRA) perspective, this construction  breaks supersymmetry explicitly rather than spontaneously. Furthermore as argued in \cite{Brustein:2004xn}, the logic of incorporating the non-perturbative effects implies that one should first find a classically stable string compactification (with at worst flat directions). The addition of D-bar branes vitiate this requirement, since they lead to a run-away potential for the \Ka modulus, decompactifying the internal manifold. Any phenomenology based on this model then is basically a test of this rather ad hoc uplift term, and so will have little to do with the underlying string theory. 

A subsequent model of IIB flux compactification, known as the Large Volume Scenario (LVS)\cite{Bala:2005zx}, overcomes some of the problems of the KKLT model. In particular while the explicit minimum obtained there still has a negative CC, it breaks SUSY. Furthermore it can be argued that the phenomenological consequences (soft masses, etc.) are not strongly affected by the mechanism by which the CC is ultimately uplifted to positive values\cite{Conlon:2008wa}\cite{Blumenhagen:2009gk}\cite{deAlwis:2009fn}\cite{Baer:2010uy}\cite{Baer:2010rz}. In LVS, the compact volume is a so called Swiss Cheese manifold, with one large \Ka modulus and one (or more) smaller \Ka moduli\footnote{The standard model fields are located on a stack of D7-branes wrapping an additional cycle which in some models tends to shrink below the string scale, or on a stack of D3 branes located at a singularity. We will assume for the purposes of this paper that the latter is the case here and will ignore this additional cycle and questions associated with its stabilization.}.  All of the moduli fields are again stabilized with a combination of fluxes and non-perturbative effects. However, this model is, in principle, susceptible to violations of constraints on flavor changing neutral currents (FCNC)\cite{deAlwis:2009fn}.  This potential violation can be traced back to fact that the model uses more than one \Ka modulus.

Essentially, the general expression for the soft masses in this model contains two terms, one flavor diagonal term coming from the large \Ka modulus, ($T_l$, $\Re(T_l)\equiv t_l$), and one flavor non-diagonal term coming from the small \Ka modulus, ($T_s$, $\Re(T_s)\equiv t_s$).  The ratio of these two terms is proportional to the ratio of their associated harmonic $(1,1)$ forms $\omega_{l}$, $\omega_{s}$ ($\omega_{l}$ dual to $t_l$, $\omega_{s}$ dual to $t_s$).  FCNC suppression then demands that $\omega_{s} \lesssim 10^{-3}\frac{1}{\ln(m_{3/2}) t_b}\omega_{l}$.  This can be achieved if the small \Ka modulus is chosen to be a blow up of a singularity some distance R from the stack of D3 branes and with R being larger than a certain lower bound (for details see \cite{deAlwis:2009fn}).
     
While, in principle, there is no problem achieving this within the LVS construction it is still worthwhile examining whether this additional input discussed above can be avoided. This leads us to examine models that use a single \Ka modulus.  We may follow the procedure of \cite{Bala:2005zx} and look for minima of the scalar potential in which the complex structure moduli are stabilized at points which are such that the SUSY breaking direction is orthogonal to these moduli.  From here, we have the choice of assuming that the axio-dilaton is also stabilized at such a point or that it contributes to the breaking of supersymmetry. \footnote{It should be noted that this procedure is just a slight extension of that followed in the original LVS paper \cite{Bala:2005zx}. Also we would like to stress that this LVS procedure is not the same as the so-called two stage procedure in which the dilaton and complex structure moduli are first integrated out (assuming that the relevant masses are high, and then studying the resulting theory for the light moduli). For some discussion on the validity of the latter see for instance \cite{deAlwis:2005tf}\cite{Gallego:2008qi}\cite{Gallego:2009px}\cite{Gallego:2011jm}\cite{Brizi:2009nn}.
}

Our strategy is to consider various SUGRA models coming from IIB flux compactification.  These models are defined by their \Ka potentials and superpotentials.  We stabilize the moduli fields in these models either analytically or numerically and we examine the relevant particle phenomenology in each case.  For the numerical results, we use standard minimization functions in Mathematica to locate minima and to evaluate the scalar potential and other quantities.  In addition, we use the program STRINGVACUA\cite{Gray:2008zs} in order to simplify these calculations but we do not make use of this program's algebraic geometry-based algorithms. We find that it is possible to find minima where supersymmetry is broken and with the scale of the cosmological constant  being close to zero. In the simplest case the gravitino and hence soft mass scale is far above the TeV scale. Hence these models, while appearing to be consistent outcomes of type IIB string theory compactified on CY orientifolds with just one \Ka modulus, do not address the hierarchy problem and hence are not relevant for physics at the LHC. Nevertheless these are simple examples of SUSY breaking models with nearly zero cosmological constant coming from string theory. To get models with TeV scale gravitino mass on the other hand requires rather complicated models with several non-perturbative terms. These we analyze numerically and we present an example with 10TeV gravitino mass.

This paper is outlined as follows.  In section 2, we investigate a simple SUGRA model in which supersymmetry is broken by the \Ka modulus using non-perturbative and $\alpha'$ corrections.  We derive both analytic and numerical results for this model.  In addition, we discuss its phenomenology.  In section 3, we derive similar results for a model in which supersymmetry is broken by both the \Ka modulus as well as the axio-dilaton.  In section 4, we summarize our results.  We conclude by examining a natural extension of our first model in the appendix. 

\section{Single \Ka Modulus + $\alpha'$ + Non-Perturbative Term}

We begin by examining a model of supergravity coming from IIB string compactifications on Calabi Yau orientifolds with D branes and fluxes\footnote{This particular model was first studied in \cite{Balasubramanian:2004uy} and \cite{Westphal:2006tn}.  We extend the study of this model by including various analytic and numerical results.}.  We assume that the MSSM lives on a stack of D3 branes at a singularity.  We consider a model with a single \Ka modulus, $T$, and an axio-dilaton, $S$, but with many complex structure moduli, $U^i$, ($i = 1,\dots,h_{21}$; $h_{21}>1$).  In addition, we include an $\alpha'$ correction \cite{Becker:2002nn} and a non-perturbative term coming from either gaugino condensation or instantons.  This model is defined by its \Ka and superpotentials given below 
\beq\label{KART}
K = -2\ln\Big( \big(\frac{1}{2}(T + \overline{T})\big)^{3/2} + \frac{\hat{\xi}}{2}\big(\frac{1}{2}(S + \overline{S})\big)^{3/2} \Big) - \ln(S + \overline{S}) -\ln(k(U,\overline{U})), 
\eeq  
\beq\label{KARTW}
W = W_{flux}(S,U) + Ae^{-aT}. 
\eeq
Here\footnote{Our notation differs slightly from \cite{Bala:2005zx}, $\hat{\xi}$ and $\xi$ are interchanged.} $\hat{\xi} = \frac{-\chi\zeta(3)}{2(2\pi)^3}$, $\chi = 2(h_{11}\!-\!h_{21})$, $U$ represents all of the $U^i$ and $a=\frac{2\pi}{N}$, where $N$ is the rank of the hidden sector gauge group. Note that since the compactifications that we consider all have $h_{21}>h_{11}$ the parameter $\hat{\xi}$ is positive. We define the complex moduli fields as $T=t+i\tau$ and $S=s+i\sigma$.  We will search for minima of this model's scalar potential that break supersymmetry along the $T$ direction.

\subsection{Analytic Results}

We begin by examining this model (eqns.~\eqref{KART},\eqref{KARTW}) analytically.  The scalar potential can be written as 
\beq\label{ScPotK}
V = e^K\left[K^{T\overline{T}}D_TWD_{\overline{T}}\overline{W} +2\Re\left(K^{S\overline{T}}D_SWD_{\overline{T}}\overline{W}\right) -3|W|^2 \right] +|F^S|^2 + |F^U|^2
\eeq  
We follow the approach of the LVS model and look for minima that break supersymmetry in a self-consistent large volume approximation
\beq\label{ToverS}
\mathcal{V}|_{min}=t^{3/2}|_{min} \gg \xi \equiv \hat{\xi} \; s^{3/2}|_{min}
\eeq
This allows us to approximate the \Ka potential and its derivative as 
\beq\label{Kapp}
K_T = K_{\overline{T}} \approx \frac{-3}{2t}\left(1-\frac{\xi}{2t^{3/2}}\right)\;\;\;\;\;\;K^{T\overline{T}} \approx \frac{4t^2}{3}\left(1+\frac{\xi}{2t^{3/2}}\right)
\eeq 
\beq\label{Eapp}
e^K = \frac{1}{\left(t^{3/2} + \frac{\xi}{2}\right)^2k(U,\overline{U})(2s)} \approx \frac{1}{t^3k(U,\overline{U})(2s)}\left(1 - \frac{\xi}{t^{3/2}}\right)
\eeq
Combining these terms together we get for the scalar potential
\begin{eqnarray}\label{PotKT}
V &\sim& \frac{1}{t^3k(U,\overline{U})(2s)}\left[\frac{4t^2}{3}\left(a^2|A|^2e^{-2at}\right)+2\Re\left((-aAe^{-aT})(-2t)\overline{W}\right) +\frac{3\xi}{4t^{3/2}}|W|^2 \right] \nonumber \\
  & &  + \ord\left(\frac{e^{-2at}}{t^{5/2}},\frac{e^{-at}}{t^{7/2}},\frac{1}{t^{9/2}}\right) +2\Re(K_{S\overline{T}}F^SF^{\overline{T}}) +|F^S|^2 + |F^U|^2
\end{eqnarray}

By extremizing the scalar potential only in the $T$ direction, we will find that $V|_{min} \sim \ord (\frac{1}{\mathcal{V}^3})$. The terms in eqn.~\eqref{PotKT} that involve $F^S$ and $F^U$ can be approximated as
\begin{eqnarray}
|F^S|^2 \sim \ord\left(\frac{1}{\mathcal{V}^2}\right) \;\;\;\; |F^U|^2 \sim \ord \left(\frac{1}{\mathcal{V}^2}\right) \nonumber \\
2\Re\left(K_{S\overline{T}}F^SF^{\overline{T}}\right) \sim \ord\left(\frac{1}{t^{5/2}}\frac{1}{t^{3/2}}\frac{1}{t^{1/2}}\right) \sim \ord\left(\frac{1}{\mathcal{V}^3}\right) 
\end{eqnarray}
Since $|F^S|$ and $|F^T|$ are both positive definite, we see that  a large volume minimum with $F^S|_{min}= F^U|_{min}=0$ obtained by looking at the $T$ minimization conditions will in fact be  a minimum of the full potential $V(S,T,U)$ because motion along any of the moduli fields away from the minimum  necessarily increases $V(S,T,U)$.

We now proceed to look at the conditions for a minimum with respect to $T$ of $V$ \footnote{This is essentially the same procedure as in  \cite{Bala:2005zx}.}. From eqn.~\eqref{PotKT} we may extract the axion dependence of the scalar potential
\begin{equation}
V(\tau) \sim \frac{1}{t^3k(U,\overline{U})(2s)}\left(2\Re\left(-aAe^{-aT}\overline{W}_0(-2t))\right)\right)
\end{equation}
We define the complex quantities as follows, $A = |A|e^{i\phi_A}$, $W_0 = |W_0|e^{i\phi_{W_0}}$ ($W_0 \equiv W(S,U)_{\text{flux}}|_{\text{min}}$).  The potential's axion dependence now becomes 
\begin{equation}\label{VTauexp}
V(\tau) \sim \frac{4ae^{-at}}{t^2}\left(|A||W_0|\cos(a\tau -\phi_{A}+\phi_{W_0})\right) 
\end{equation}
Where we have assumed that $\frac{1}{(2s)(k(U,\overline{U}))}|_{min}\sim\ord(1)$. Extremizing with respect to $\tau$,
\begin{equation}\label{VK_ax_ext}
V'(\tau) = \frac{-4a^2e^{-at}}{t^2}\left(|A||W_0|\sin(a\tau-\phi_{A}+\phi_{W_0})\right) = 0
\end{equation}  
The set of solutions to this equation is
\beq\label{VK_min_pnts}
a\tau-\phi_A+\phi_{W_0} = n\pi \;\;\;\;\; n \in \mathbb{Z}
\eeq

This set of solutions gives us insight into the structure of the Hessian matrix.  In order to find minima of the potential, we must find extrema for which the eigenvalues of the Hessian matrix are all positive.  From eqn.~\eqref{VK_ax_ext} and \eqref{VK_min_pnts} we see that the off-diagonal terms vanish, $\frac{\partial^2V}{\partial \tau \partial t}|_{min} = \frac{\partial^2V}{\partial t \partial \tau}|_{min} = 0$.  This simplifies the Hessian matrix to the following form
\[
\left( 
\begin{array}{cc}
\frac{\displaystyle\partial^2V}{\displaystyle\partial t^2} &\displaystyle 0  \\
\displaystyle 0 & \frac{\displaystyle\partial^2V}{\displaystyle\partial\tau^2}
\end{array} 
\right)        
\]
From this matrix, we see that both eigenvalues are positive if and only if both $\partial^2_tV$ and $\partial^2_{\tau}V$ are also positive.  

We now check the concavity of the potential at the $\tau$ extremum,
\begin{equation}
V^{\prime\prime}(\tau) = \frac{-4a^3e^{-at}}{t^2}\Big(|A||W_0|\cos(a\tau-\phi_{A}+\phi_{W_0})\Big)
\end{equation}
In order to isolate a minimum, we require $V^{\prime\prime}>0$, therefore
\beq\label{KSRT_axion}
a\tau-\phi_A+\phi_{W_0} = (2n+1)\pi \;\;\;\; n \in \mathbb{Z}
\eeq

Inserting eqn.~\eqref{KSRT_axion} into eqn.~\eqref{VTauexp} with $F^S=F^U=0$, we compute the scalar potential for this model and expand in negative powers of the volume. 
For large volumes  the potential can be safely approximated by
\begin{equation}\label{Vapprox}
V \sim \frac{4}{3}\Big(a^2|A|^2e^{-2at}\Big)\frac{t^{1/2}}{\mathcal{V}}+4\Big(a|A|^2e^{-2at} -a|W_0||A|e^{-at} \Big)\frac{t}{\mathcal{V}^2} + \frac{3|W_0|^2\xi}{4\mathcal{V}^3} + \dots 
\end{equation}
Where we have again assumed that $\frac{1}{(2s)(k(U,\overline{U}))}|_{min}\sim\ord(1)$.

From here, the scalar potential can be further simplified with knowledge of the magnitude of $W_0$.  There are two relevant regimes, $|W_0| \sim e^{-at}$ and $|W_0| \gg e^{-at}$ that may lead to the sort of minimum we are looking for.  In the first regime we see that the $\alpha '$ correction term (the last term of eqn.~\eqref{Vapprox}) can be ignored. This is then essentially the KKLT situation and the corresponding minimum is supersymmetric. The numerical search for minima in this limit confirm that  such minima are indeed supersymmetric.  

We now investigate the remaining regime, $|W_0| \!\gg\! e^{-at}$.  In this limit, the scalar potential is exponentially suppressed at large volumes and simplifies to 
\begin{equation}\label{KSRT_pot}
V \sim -\Big(4 |W_0|(a|A|e^{-at})\Big)\frac{t}{\mathcal{V}^2} + \frac{3W_0^2\xi}{4\mathcal{V}^3} + \dots 
\end{equation}
We solve for the minimum of this potential by suppressing the term in the derivative that is $\sim\ord\left(\frac{W_0e^{-at}}{t^3}\right)$.  This is tantamount to assuming that $at \gtrsim \ord(2)$.  The extremization condition ($\partial_tV = 0$) yields the relation
\beq\label{W0_min}
|W_0| =  \frac{32}{27\xi}\left(a^2|A|e^{-at}\right)t^{7/2}  
\eeq
This shows that at the minimum of the potential, $|W_0|$ is much larger than $e^{-at}$, which is consistent with our original assumption. Checking for positive concavity of the minimum and using the same approximation ($at \gtrsim \ord(2)$) gives the condition
\beq\label{ATBnd}
V'' = \frac{27|W_0|^2\xi}{8t^{11/2}}\Big(-a + \frac{11}{2t}\Big)  > 0
\eeq
We see from this equation that for $at\!<\!11/2$ this extremum is a minimum (note that this is essentially a condition relating the fluxes and $a$  as is evident from eqn.~\eqref{W0_min}).  Therefore, the gravitino mass is bounded from below by
\beq
m_{3/2} \sim \frac{|W_0|}{t^{3/2}} \gtrsim \frac{e^{-11/2}\left(\frac{11}{2}\right)^2}{\xi} \sim 10^{-3}\;\text{M}_{\text{P}}\;\; \text{or}\;\; 10^{15} \; \text{GeV} 
\eeq
For $\xi\sim \ord(100)$.
We may estimate the value of the scalar potential at the minimum by inserting eqn.~\eqref{W0_min} into eqn.~\eqref{KSRT_pot}.  This yields the following relation
\begin{eqnarray}\label{KSRT_Vmin}
V|_{min} &=& -\Big(4|W_0|\Big(+\frac{27\xi |W_0|}{32t^{7/2}a} \Big)\Big)t^{-2} +\frac{3|W_0|^2\xi}{4t^{9/2}} \nonumber \\
 &=& \frac{3|W_0|^2\xi}{4t^{9/2}}\Big(-\frac{9}{2at}+1\Big) 
\end{eqnarray} 
For $at \sim \ord (1)$, $V|_{min} \sim \ord \left(\frac{1}{\mathcal{V}^3}\right)$.  This is in agreement with our original assertion about the scale of $V|_{min}$, namely, for large volumes, $V|_{min}$ is suppressed relative to the terms in the full potential $V(T,S,U)$ that are proportional to $F^S$ or $F^U$.  Therefore, this is a minimum of the full scalar potential.  It is a deSitter minimum for $\frac{9}{2}<at<\frac{11}{2}$.  It is important to reiterate that this bound on $at$ is approximate and principally used to make an order of magnitude estimate on the lower bound of $m_{3/2}$.  Exact bounds on $at$ necessary for a deSitter minimum require one to numerically solve\footnote{We thank Alexander Westphal and Markus Rummel for discussing this issue.  The explicit calculation of the deSitter bounds of $at$ is performed in \cite{Rummel:2011cd}.} the conditions $\partial_tV = 0$ and $\partial^2_tV > 0$.

We may check the stability of this minimum against the well known necessary criteria established in the work of Covi et.al.\cite{Covi:2008ea} (see eq. 5.35) as well as \cite{GomezReino:2006dk}.  The relevant bound is
\beq
\tilde{\delta}\equiv \frac{\xi}{16\mathcal{V}} \geq \frac{2V|_{min}}{105m_{3/2}^2}
\eeq
For our model, we maximize $V|_{min}$ and observe that
\beq
\frac{\xi}{16\mathcal{V}} \geq \frac{2\!\times\!2\!\times\! 3 |W_0|^2\xi}{105\!\times\!11\!\times\!4t^{9/2}m_{3/2}^2} = \frac{\xi}{385\mathcal{V}}
\eeq
Therefore, we confirm that this necessary condition is indeed satisfied.

We may also check whether this minimum is stable under quantum corrections.  As discussed in \cite{Berg:2005ja}\cite{Berg:2005yu}\cite{vonGersdorff:2005bf}, the \Ka potential (eqn.~\eqref{KART}) receives corrections at 1-loop of the form
\beq
K\rightarrow K + \frac{1}{T+\overline{T}}\left[\frac{f(A,\overline{A},U,\overline{U})}{S +\overline{S}}\right] + \dots
\eeq 
Here, $f(A,\overline{A},U,\overline{U})$ is a function of the open string scalars as defined in \cite{Berg:2005yu}. For our model, this translates into a scalar potential of the form
\beq\label{Vmin_Loop}
V = \left[\frac{c_1(S+\overline{S})^{3/2}}{(T+\overline{T})^{9/2}} + \frac{c_2}{(T+\overline{T})^{10/2}(S+\overline{S})^2} + \frac{c_3(S+\overline{S})^{3/2}}{(T+\overline{T})^{11/2}} +\dots\right]|W_0|^2
\eeq
where $c_i\lesssim \ord (10)$.  Comparing this with eqn.~\eqref{KSRT_Vmin}, we may identify the 1-loop correction as the term $\sim\ord\left(\frac{1}{(T+\overline{T})^{10/2}}\right)$.  We see that for $s\sim\ord(1)$ the 1-loop correction indeed alters our minimum.  In order to suppress this correction we need to choose fluxes such that the value of $s$ is large enough.
From eqn.~\eqref{Vmin_Loop}, we find that for 
\beq
(S+\overline{S})\gtrsim(T+\overline{T})^{1/7}
\eeq    
the quantum term in eqn.~\eqref{Vmin_Loop} can be ignored and we recover our original minimum.  For example, if $t\sim 10$, $s$ must be $\gtrsim 1.4$ to suppress the quantum correction\footnote{Consistency of the two super-covariant derivative expansion when the lightest integrated-out scale is the Kaluza-Klein scale requires $|W_0| < t^{-1/2}$.  This implies $t \lesssim \ord(10)$.}. 

We now calculate the classical soft masses using the general expression \cite{Kaplunovsky:1993rd}\cite{Brignole:1997dp}
\beq\label{SoftMass}
m^2_{\alpha\overline{\beta}} = V|_{min}K_{\alpha\overline{\beta}} + m^2_{3/2}K_{\alpha\overline{\beta}} - F^AF^{\overline{B}}R_{A\overline{B}\alpha\overline{\beta}}
\eeq
For our model this reduces to
\beq
m^2_{\alpha\overline{\beta}} \sim m^2_{3/2}K_{\alpha\overline{\beta}} -F^TF^{\overline{T}}R_{T\overline{T}\alpha\overline{\beta}}
\eeq
The calculation of the Riemann curvature tensor and the F-terms may be adapted from the results derived in \cite{deAlwis:2009fn} which follow from \cite{Grana:2003ek} and \cite{Covi:2008ea}.  We quote the value of the soft mass, $m^2_s$, (where $m^2_{\alpha\overline{\beta}}\equiv m^2_s K_{\alpha\overline{\beta}}$) below
\beq
m_s^2 = \frac{5\xi}{8t^{3/2}}m^2_{3/2}
\eeq
We conclude that the soft masses are not tachyonic (since $\xi$ is positive).  However, they are fixed at a scale comparable to $m_{3/2}$, i.e.\! parametrically above the weak scale and are thus of limited phenomenological interest. 

\section{Single \Ka Modulus with $S$ and $T$ SUSY breaking}
\subsection{Series Expansion Analysis}
We now investigate a class of SUGRA models in which supersymmetry can be broken in both the $S$ and $T$ directions.  As in the previous example, we study models coming from IIB compactifications on Calabi Yau orientifolds with matter living on D3 branes at a singularity.  We include Wilson lines in the compactification in order the break the gauge group into a direct product group $\Pi_i SU(N_i)$.  We assume that these groups condense to give non-perturbative corrections to the superpotential that break supersymmetry.  Unlike the previous model, we do not include $\alpha^{\prime}$ corrections to the \Ka potential.  The generic expressions for the \Ka and superpotentials are given below
\beq
K = -3\ln(T + \overline{T}) -\ln(S + \overline{S}) -\ln(k(U,\overline{U}))
\eeq  
\beq\label{W_Full}
W = A(U) + B(U)S + \sum_i C_i(U,S)e^{-x_iT}
\eeq
Here, $x_i \equiv \frac{2\pi}{N_i}$ where $N_i$ is the rank of the $i\text{th}$ gauge group and $U$ represents all of the complex structure moduli $(U^a , a = 1,\dots,h_{21})$.  For our analysis we will assume that the exponential prefactors $C_i$ are $\ord(1)$ and that their $U$ and $S$ dependence comes from threshold effects and internal fluxes, i.e. $C_i(U,S) = C_i(U)e^{\alpha_iS}$. (See for example \cite{Blumenhagen:2008aw}\footnote{In this paper the fluxes are used to break the $SU(5)$ gauge group containing the standard model. Here by contrast we are breaking the condensing group which generates the non-perturbative terms in $W$.}) Therefore, the superpotential can be written as
\beq
W = A(U) + B(U)S + \sum_i C_i(U)e^{-x_iT + \alpha_i S}
\eeq
 
Let us now examine a technique for handling this model numerically\footnote{The following method was first outlined in \cite{Brustein:2004xn}.}. Suppose that we identify a minimum of the scalar potential at a point, $(S_0, T_0, U_0)$ in field space.  Without loss of generality we assume that this point is real.  We expand the superpotential only in fluctuations about the $S$ and $T$ directions. We assume that there is sufficient freedom in the choice of fluxes that once the minimization in these two directions are carried out fluxes can be chosen such that this remains a minimum with some value of $U$ such that $F^U=0$. With a sufficient number of 3-cycles this should be always possible. We expand $W$ as
\beq
W(S,T,U) = \sum_{n,m} a_{nm}(U)(S-S_0)^n(T-T_0)^m
\eeq
Comparing this with eqn.~\eqref{W_Full} gives
\begin{eqnarray}\label{W_exp}
a_{nm} & = & \frac{1}{n!m!}\partial^n_S\partial^m_TW_0 \nonumber \\
       & = & \frac{1}{n!m!}[(A_0 + S_0B_0)\delta_{n0}\delta_{m0} + B_0\delta_{n1}\delta_{m0} + \sum_i(-x_i)^m\partial^n_SC_{i0}e^{-x_iT_0}] \nonumber \\
       & \equiv & e^{-x_iT_0}S_0^{-n}T_0^{-m}\tilde{a}_{nm}
\end{eqnarray}
Where $W_0 \equiv W(S_0,T_0,U_0)$.  We now redefine the fields as ($\widetilde{S} \equiv S/S_0$, $\widetilde{T} \equiv T/T_0$).  We may then write the superpotential as
\beq
W = e^{-x_iT_0}\sum_{nm}\tilde{a}_{nm}(\widetilde{S}-1)^n(\widetilde{T}-1)^m \equiv e^{-x_iT_0}\widetilde{W}
\eeq 
This results in an overall scaling of the scalar potential
\beq
V = \frac{e^{-2x_iT_0}}{T^3_0S_0}\widetilde{V}(\widetilde{S},\widetilde{T},U,\overline{\widetilde{S}},\overline{\widetilde{T}},\overline{U})
\eeq
where $\widetilde{V}$ is defined in terms of $\widetilde{W}$ and $\widetilde{K} = -3\ln(\widetilde{T} + \overline{\widetilde{T}}) -\ln(\widetilde{S} + \overline{\widetilde{S}}) -\ln(k(U,\overline{U}))$.

Expanding the superpotential in a Taylor series allows us to control the location and value of scalar potential's minimum.  Since the Hessian matrix for the scalar potential only depends on terms up to third order in the expanded superpotential, we can arbitrarily tune a minimum of the scalar potential by solving the following system of equations (from eqn.~\eqref{W_exp})
\begin{eqnarray}\label{W_dict}
\tilde{a}_{00} & = & e^{x_1T_0}(A_0 + S_0B_0) + \sum_{i}C_{i0}e^{-(x_i -x_1)T_0} \nonumber \\
\tilde{a}_{10} & = & S_0\big[e^{x_1T_0}B_0 + \sum_{i}\partial_SC_{i0}e^{-(x_i -x_1)T_0}\big] \nonumber \\
\tilde{a}_{01} & = & T_0 \sum_{i}(-x_i)C_{i0}e^{-(x_i -x_1)T_0}\nonumber \\
               & \vdots  & \nonumber \\
\tilde{a}_{30} & = & \frac{S_0}{6} \sum_{i}\partial_SC_{i0}e^{-(x_i -x_1)T_0} 
\end{eqnarray}

\subsection{Numerical Example}
Following the arguments of the previous section we consider the following SUGRA model
\beq\label{KFull}
K = -3 \ln(T+\overline{T}) - \ln(S+\overline{S}) -\ln(k(U,\overline{U}))
\eeq
\beq\label{WFull}
W = A_0 + B_0*S + C_1e^{-x_1T + \alpha_1 S} + C_2e^{-x_2T + \alpha_2 S} + C_3e^{-x_3T + \alpha_3 S} + C_4e^{-x_4T + \alpha_4 S}
\eeq
We include four non-perturbative terms because expanding the superpotential to third order requires ten independent parameters.  If we want to construct a minimum of the scalar potential with the gravitino mass fixed to a certain scale it turns out that  unless we include four non-perturbative terms it is too hard to solve for a minimum.  We can construct an extremum of the scalar potential with two or three non-perturbative terms but we cannot guarantee that such an extremum is a minimum because we lack enough free parameters to simultaneously solve all ten equations given above (eqn.~\eqref{W_dict}).  

For models with two or three non-perturbative terms, requiring the extremum to be a minimum, in principle, defines some region in 3-dimensional parameter space.  (e.g.\ $\{(\tilde{\alpha}_{00},\tilde{\alpha}_{10},\tilde{\alpha}_{01})\}$).  This region is identified by requiring the eigenvalues of the Hessian to be positive definite.  However, general expressions for the eigenvalues are complicated enough to prevent the identification of this region in a computationally tractable manner.  Therefore, including four non-perturbative terms and solving the system of equations given above (eqn.~\eqref{W_dict}) is the most reliable technique for identifying a minimum in this class of models.
 
\begin{figure}
\centering
\includegraphics[scale=0.75]{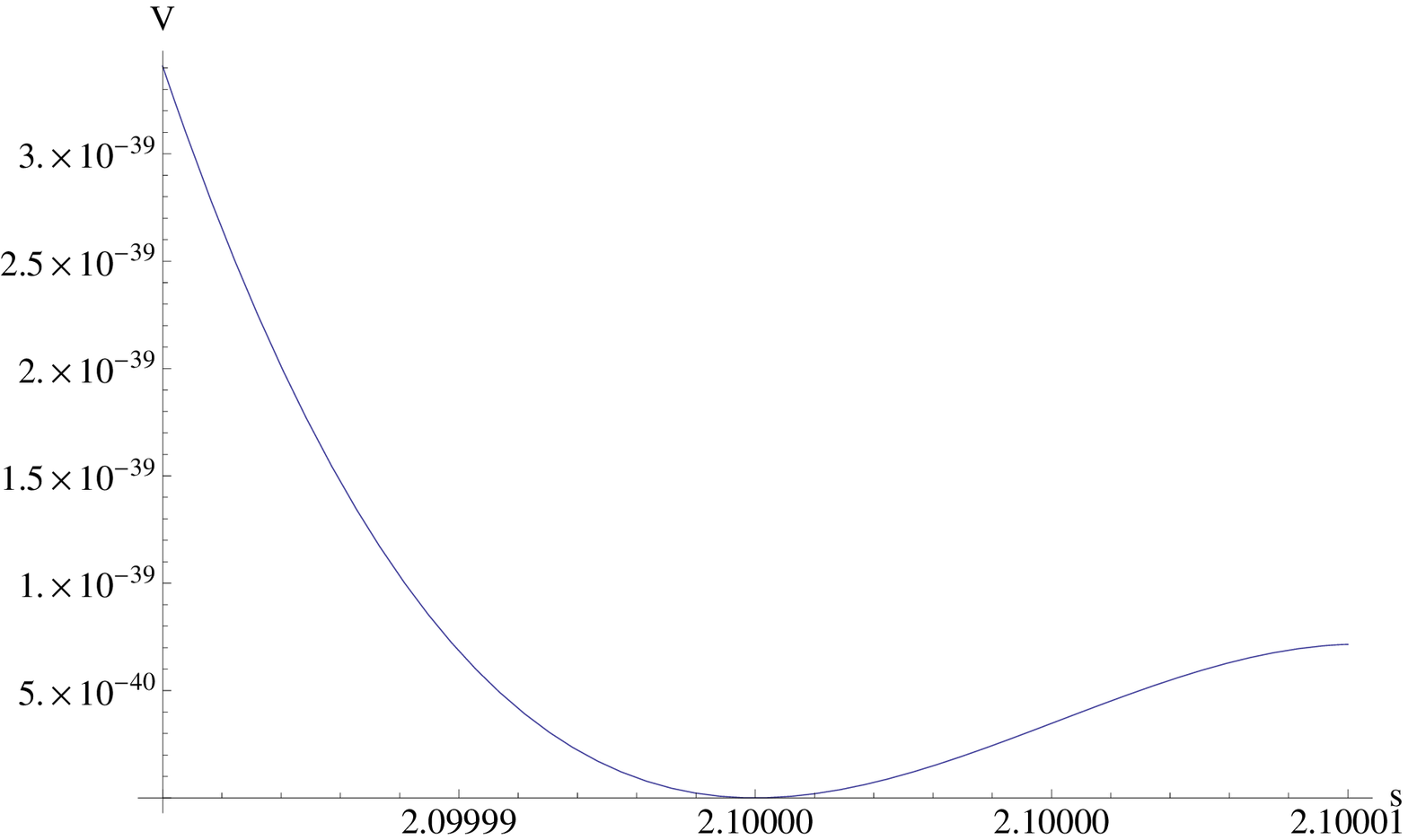}
\caption{Vmin for $<\!\!t\!\!> =40$}
\label{Sfull}
\end{figure}

\begin{figure}
\centering
\includegraphics[scale=0.75]{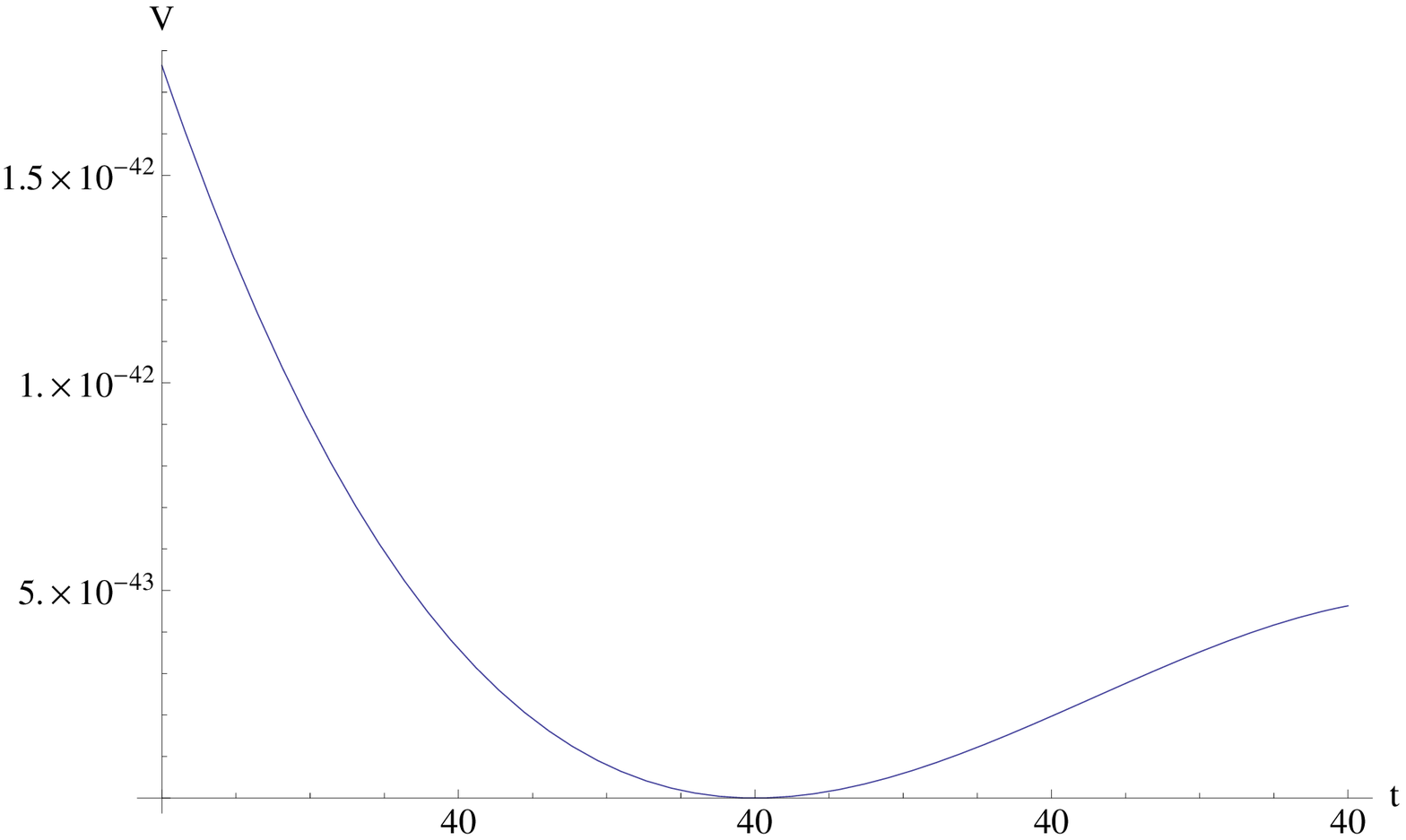}
\caption{Vmin for $<\!\!s\!\!> =2.1$}
\label{Tfull}
\end{figure}

From these arguments we construct a Minkowski minimum with  $m_{3/2}\sim 10 \;\text{TeV}$ for the following values of the parameters given in table~\ref{T4}. Plots of this minimum along the s, ($\Re(S)$), and t, ($\Re(T)$), directions are given in figures~\ref{Sfull} and \ref{Tfull}.  This minimum is adapted from the local model identified in \cite{Brustein:2004xn}.

{
\renewcommand{\arraystretch}{1.3}
\begin{table}[h]
\begin{center}
\begin{tabular}{|l|l|}
  \hline
   & $A_0 =1.85*10^{-8}$\;\;\; $B_0 = 1.6*10^{-10}$\;\;\;$C_1 = -3.4$\;\;\; $x_1 = \frac{2\pi}{30}$\\
   & $\alpha_1 = -1.06$ \;\;$C_2 = 13.3$\;\; $x_2 = \frac{2\pi}{29}$\;\; $\alpha_2 = -1.1$ \;\; $C_3 = -17.7$\\
   & $x_3 = \frac{2\pi}{28}$\;\;\; $\alpha_3= -1.14$\;\;\; $C_4= 8.1$\;\;\;$x_4=\frac{2\pi}{27}$\;\;\;$\alpha_4= -1.18$\\[2pt]
  \hline
  $<\!\!t\!>$ & $40$ \\
  $<\!\!\tau\!>$ & $0$ \\
  $<\!\!s\!>$ & $2.1$ \\
  $<\!\!\sigma\!>$ & $0$ \\
  $V_0|_{min}$ & $0$  \\
  $m_{3/2}^2\equiv e^K|W|^2$ & $1.3\!\times\!10^{-28}$  \\
  $|F^T|^2K_{T\overline{T}}\equiv e^KK^{T\overline{T}}|D_TW|^2$ &  $3.2\!\times\!10^{-28}$  \\
  $|F^S|^2K_{S\overline{S}}\equiv e^KK^{S\overline{S}}|D_SW|^2$ &  $6.8\!\times\!10^{-29}$  \\
  \hline
\end{tabular}
\caption{Moduli field vev's, F-terms, Gravitino mass and the Cosmological Constant for SKM + 4 Non-Pert Terms at a non-SUSY minimum of the scalar potential (in $M_P=1$ units).}
\label{T4}
\end{center}
\end{table}
}

In this example we note that, at the minimum of the potential, $|F_S|\sim 4|F_T|$.  In principle we expect $|F_S|$ and $|F_T|$ to be of the same order.  In fact, the relatively low scale of $m_{3/2}$ for this model depends on these two F-terms making comparable contributions to the SUSY breaking. In the limit of $|F_S| \!\rightarrow\! 0$ with $|F_T| \ne 0$ we return to the situation described by well-known no-go theorems \cite{GomezReino:2006dk}\cite{Covi:2008ea}\cite{Brustein:2004xn} and there would be no deSitter minimum. When $|F_S|$ is non-zero but subdominant to $|F_T|$ we may plausibly recover a high scale deSitter minimum, analogous to the previous model, with the axio-dilaton playing the role of a subdominant correction to the \Ka modulus.  In either case, a low scale deSitter minimum depends crucially on that fact that $|F_S| \sim |F_T|$. 
%The fact that, for this example, $|F_S| \ne 1*|F_T|$ is seen as an artifact of the particular choice of parameters and not endemic to the class of minima.   

We may calculate the soft masses for this particular example by following the approach of \cite{deAlwis:2008kt}. Namely, we may express the full \Ka potential, including matter fields as
\beq
K = K_{mod} + Z(T)_{\alpha\overline{\beta}}\Phi^{\alpha}\overline{\Phi}^{\overline{\beta}} + \dots
\eeq
Where, $Z(T)_{\alpha\overline{\beta}} = \frac{3\delta_{\alpha\overline{\beta}}}{T+\overline{T}}$ and $K_{mod}= -3\ln(T+\overline{T})-\ln(S+\overline{S})-\ln(k(U,\overline{U}))$. The soft masses can be calculated from the \Ka potential following the general expression given in eqn.~\eqref{SoftMass}.  The only relevant non-vanishing curvature component is $R_{T\overline{T}\alpha\overline{\beta}} = \frac{1}{3}K_{T\overline{T}}Z_{\alpha\overline{\beta}} + \ord(\Phi^2)$.  Therefore, for this model, the soft mass expression becomes
\beq
m_s^2 Z_{\alpha\overline{\beta}} = \left(m_{3/2}^2 - \frac{1}{3}F^T\overline{F}^{\overline{T}}K_{T\overline{T}}\right)Z_{\alpha\overline{\beta}} = \frac{1}{3}F^S\overline{F}^{\overline{S}}K_{S\overline{S}} Z_{\alpha\overline{\beta}}
\eeq
Therefore, $m_{s}^2 \!\approx\! 2.2\times10^{-29} \;M_P$ or $m_s \!\approx\! 4.8 \;\text{TeV}$.  Note that as long as $V_0 \!\ll\! m_{3/2}^2$ for this class of models, $m_{s}^2$ will always be roughly equal to $\frac{1}{3}|F^S|^2$ and hence positive.

It is worth reiterating that this specific model, including all its relevant scales, has been arbitrarily chosen.  We are free, in principle, to generate a model with any desired scale by solving the corresponding system of equations (eqn.~\eqref{W_dict}).  What we have demonstrated is a general technique for finding such models. 

\section{Conclusion}

We have demonstrated that there exists physically plausible vacua coming from IIB string compactifications on Calabi-Yau orientifolds having one \Ka modulus together with fluxes and D-Branes.  Such models have natural FCNC suppression due to the fact that they contain only one \Ka modulus\footnote{Quantum corrections will not alter this picture due to the large volume suppression, see \cite{Burgess:2010sy}.}. In the simplest model, (eqns.~\eqref{KART},\eqref{KARTW}), an $\alpha'$ correction allows SUSY to be broken along the $T$, (\Ka modulus), direction.  A Minkowski or de Sitter classical minimum is attainable but the soft mass phenomenology is such that it is of no relevance for the hierarchy problem. This is due to the fact that the gravitino mass is fixed at a high scale ($m_{3/2} \gtrsim 10^{-3} \times M_P$).

In the second model, (eqns.~\eqref{KFull},\eqref{WFull}), the gravitino mass can be set to any scale by appropriate choice of fluxes.  SUSY is broken in both the $S$, (axio-dilaton), and $T$, (\Ka modulus), directions and we expect both fields to contribute comparable F-terms.  The classical cosmological constant as well as the location of the minimum in field space can be tuned by solving the appropriate equations coming from the Taylor series expansion of the superpotential (eqn.~\eqref{W_dict}). However, in order to solve these equations in a tractable manner, the superpotential must include at least four non-perturbative terms.    

Finally let us observe that while in principle it is possible to find models (as demonstrated by the above numerical example) that can in fact give a  phenomenology that is relevant to TeV scale physics, it is hard to obtain generic consequences of the entire class of such models. The phenomenology is clearly quite sensitive to the  model parameters (fluxes choices). This is quite unlike the case of LVS models where with a few general assumptions about the location of the MSSM a viable phenomenology is obtained \cite{Conlon:2008wa}\cite{Blumenhagen:2009gk}\cite{deAlwis:2009fn}\cite{Baer:2010uy}\cite{Baer:2010rz}. While the original motivation for this investigation was in fact to remove the requirement on the location  on the MSSM cycle, that is needed in the LVS case, to satisfy FCNC constraints, the upshot of our investigation actually strengthens the case for this scenario.

\section{Acknowledgements}
We thank James Gray for correspondence concerning the STRINGVACUA program. The research of SdA and KG is partially supported by the United States Department of Energy under grant DE-FG02-91-ER-40672.
\section{Appendix: Single \Ka Modulus + $\alpha'$ + RaceTrack}

We may naturally extend our first model, (eqns.~\eqref{KART},\eqref{KARTW}) to include the effects of two non-perturbative corrections to the superpotential.  This model is given below
\beq\label{AKART}
K = -2\ln\Big( \big(\frac{1}{2}(T + \overline{T})\big)^{3/2} + \frac{\hat{\xi}}{2}\big(\frac{1}{2}(S + \overline{S})\big)^{3/2} \Big) - \ln(S + \overline{S}) -\ln(k(U,\overline{U}))
\eeq
\beq\label{AKARTW}
W = W_{flux}(S,U) + Ae^{-aT} + Be^{-bT}
\eeq
Here $\hat{\xi} = \frac{-\chi\zeta(3)}{2(2\pi)^3}$, $\chi = 2(h_{11}\!-\!h_{21})$ and $a=\frac{2\pi}{N}$, $b=\frac{2\pi}{M}$, where $N$ and $M$ are the ranks of two hidden sector gauge groups.  We may naively believe that is model will yield an improvement on the first model, but as we shall see, this improvement is only minor. Ultimately, the gravitino mass is still fixed near the Planck scale. As before we define the complex moduli fields as $T=t+i\tau$ and $S=s+i\sigma$ and we search for minima of this model's scalar potential that break supersymmetry along the $T$ direction.

\subsection{Analytic Results}

As with our first model, we may identify minima of the full scalar potential, $V(S,T,U)$, by minimizing $V(T)$ with $F^S|_{min}=F^U|_{min}=0$.  Our analytic results are essentially a straight forward generalization of the simpler model.  We present them here with a modicum of redundancy.

Taking the large volume approximations (eqns.~\eqref{ToverS},\eqref{Kapp},\eqref{Eapp}) we get a full expression for the scalar potential
\begin{eqnarray}\label{APotKT}
V &\sim& \frac{1}{t^3k(U,\overline{U})(2s)}\Big[\frac{4t^2}{3}\big(a^2|A|^2e^{-at} +b^2|B|^2e^{-2bt}+2\Re\big(aAe^{-aT}b\overline{B}e^{-b\overline{T}}\big)\big) \nonumber \\
  & & + 2\Re\big((-aAe^{-aT}-bBe^{-bT})(-2t)\overline{W}\big) +\frac{3\xi}{4t^{3/2}}|W|^2 \Big] 
\end{eqnarray}
From eqn.~\eqref{APotKT} we may extract the axion dependence of the scalar potential
\begin{eqnarray}
V(\tau) &=& \frac{1}{t^3k(U,\overline{U})(2s)}\Big(2\Re\big(-aAe^{-aT}\overline{W}_0(-2t)-aAe^{-aT}\overline{B}e^{-b\overline{T}}(-2t)-bBe^{-bT}\overline{W}_0(-2t) \nonumber \\
  & & -bBe^{-bT}\overline{A}e^{-a\overline{T}}(-2t) +\frac{4t^2}{3}aAe^{-aT}b\overline{B}e^{-b\overline{T}}\big)\Big)
\end{eqnarray}
We define the complex quantities as follows, $A = |A|e^{i\phi_A}$,  $B = |B|e^{i\phi_B}$, $W_0 = |W_0|e^{i\phi_{W_0}}$, ($W_0\equiv W_{flux}|_{min}$) .  The potential's axion dependence now becomes
\begin{eqnarray}\label{AVK_tau}
V(\tau) &=& \frac{1}{t^3}\Big(4ta|A||W_0|e^{-at}\cos(a\tau -\phi_{A}+\phi_{W_0}) + 4tb|B||W_0|e^{-bt}\cos(b\tau- \phi_{B}+\phi_{W_0}) \nonumber \\
  & & (\frac{8}{3}t^2ab+4at+4bt)|A||B|e^{-(a+b)t}\cos((a-b)\tau -\phi_{A}+\phi_{B})\Big) 
\end{eqnarray}
Where we have again assumed $\frac{1}{k(U,\overline{U})(2s)}\sim\ord(1)$. Extremizing with respect to $\tau$,
\begin{eqnarray}\label{AVK_ax_ext}
V'(\tau) &=& \frac{1}{t^3}\Big(-4ta^2|A||W_0|e^{-at}\sin(a\tau-\phi_{A}+\phi_{W_0}) -4tb^2|B||W_0|e^{-bt}\sin(b\tau- \phi_{B}+\phi_{W_0}) \nonumber \\
  & & -(a-b)(\frac{8}{3}t^2ab+4at+4bt)|A||B|e^{-(a+b)t}\sin((a-b)\tau -\phi_{A}+\phi_{B})\Big) = 0
\end{eqnarray}
The only set of solutions to this equation that is independent of $|A|$,$|B|$ and $|W_0|$ is
\beq\label{AVK_min_pnts}
a\tau-\phi_A+\phi_{W_0} = n\pi \;\;\;\;\; b\tau-\phi_B+\phi_{W_0} = m\pi\;\;\;\; n,m \in \mathbb{Z}
\eeq
We now check the concavity of the potential at the $\tau$ extremum,
\begin{eqnarray}
V^{\prime\prime}(\tau) &=& \frac{1}{t^3}\Big(-4ta^3|A||W_0|e^{-at}\cos(a\tau-\phi_{A}+\phi_{W_0}) -4tb^3|B||W_0|e^{-bt}\cos(b\tau- \phi_{B}+\phi_{W_0}) \nonumber \\
  & & -(a-b)^2(\frac{8}{3}t^2ab+4at+4bt)|A||B|e^{-(a+b)t}\cos((a-b)\tau -\phi_{A}+\phi_{B})\Big)
\end{eqnarray}

In order to isolate a minimum, we require $V^{\prime\prime}>0$.  This condition, in turn, depends on the value of $t$, $a$, $b$, $|A|$, $|B|$ and $|W_0|$.  In the limit where $|W_0| \gg e^{-at}$, $V^{\prime\prime}$ can be made positive if
\beq\label{AKSRT_axion}
a\tau-\phi_A+\phi_{W_0} = (2n+1)\pi \;\;\;\;\; b\tau-\phi_B+\phi_{W_0} = (2m+1)\pi\;\;\;\; n,m \in \mathbb{Z}
\eeq

Inserting eqn.~\eqref{AKSRT_axion} into eqn.~\eqref{AVK_tau}, we compute the scalar potential for this model and expand in negative powers of the volume.  For large volumes the potential can be safely approximated by
\begin{eqnarray}
V &\sim& \frac{4}{3}\Big(b^2|B|^2e^{-2bt} + a^2|A|^2e^{-2at} +2ab|A||B|e^{-(a+b)t} \Big)\frac{t^{1/2}}{\mathcal{V}} \\ 
& & + 4\Big(b|B|^2e^{-2bt} + a|A|^2e^{-2at} +|A||B|(a+b)e^{-(a+b)t} -|W_0|(a|A|e^{-at} +b|B|e^{-bt}) \Big)\frac{t}{\mathcal{V}^2} \nonumber \\ 
& & + \frac{3|W_0|^2\xi}{4\mathcal{V}^3} + \dots  \nonumber
\end{eqnarray}
From here, the scalar potential can be further simplified with knowledge of the magnitude of $W_0$.  Again, there are two relevant regimes; assuming $a \sim b$, $|W_0| \!\sim\! e^{-at}$ and $|W_0| \!\gg\! e^{-at}$.  As in the simpler model, minima in the first regime ($a \sim b$, $|W_0| \!\sim\! e^{-at}$) are supersymmetric.    
%We first consider $|W_0|\!\sim\! e^{-at}$.  
One may see this by examining the potential in this regime. With the benefit of foresight, we first assume that $|W_0| \approx (at) e^{-at}$.  In this limit, the scalar potential is volume suppressed yielding
\begin{eqnarray}
V &\sim& \frac{4}{3}\Big(b^2|B|^2e^{-2bt} + a^2|A|^2e^{-2at} +2ab|A||B|e^{-(a+b)t} \Big)\frac{t^{1/2}}{\mathcal{V}} \\ 
& & + 4\Big(-|W_0|(a|A|e^{-at} +b|B|e^{-bt})  \Big)\frac{t}{\mathcal{V}^2} \nonumber
\end{eqnarray}
One can solve for the minimum of the scalar potential. At this minimum, $|W_0|$ is
\beq
|W_0| = \frac{2}{3}\left(\frac{b^3|B|^2e^{-2bt} + a^3|A|^2e^{-2at} +(a+b)ab|A||B|e^{-(a+b)t}}{a^2|A|e^{-at} +b^2|B|e^{-bt}}\right)\;t \sim \mathcal{O}\left((at)e^{-at}\right)
\eeq
This is consistent with our original assumption, $|W_0| \!\approx\! (at)e^{-at}$. As in the simpler model, this minimum is supersymmetric.  One can see this by examining the F-term flatness equation.
\beq
D_T W = \partial_T W + K_T W = -aAe^{-aT} -bBe^{-bT} -\frac{3t^{1/2}W}{2t^{3/2}+ \xi} = 0
\eeq
Therefore, at the minimum,
\beq
|W_0| \sim (at)e^{-at}
\eeq
This is the same order of magnitude estimate that we initially assumed. The numerical search for minima in this limit confirm that all such minima are indeed supersymmetric.  

We now investigate the remaining regime, $|W_0| \!\gg\! e^{-at}$.  In this limit, the scalar potential is exponentially suppressed at large volumes and simplifies to 
\begin{equation}\label{AKSRT_pot}
V \sim -\Big(4 |W_0|(a|A|e^{-at} +b|B|e^{-bt})\Big)\frac{t}{\mathcal{V}^2} + \frac{3W_0^2\xi}{4\mathcal{V}^3} + \dots 
\end{equation}
Solving for the minimum and assuming that $at \sim bt \gtrsim \ord (2)$ (as in the earlier model) gives the condition
\beq\label{AW0_min}
|W_0| =  \frac{32}{27\xi}\Big((a^2|A|e^{-at} +b^2|B|e^{-bt}) \Big)t^{7/2}  
\eeq
This shows that at the minimum of the potential, $|W_0| \!\gg\! e^{-at}$, which is consistent with our original assumption. Checking for positive concavity of the minimum gives
\beq\label{AATBnd}
V'' = \frac{27|W_0|^2\xi}{8t^{11/2}}\left(-a + \frac{11}{2t}\right) -\frac{4|W_0|b^2}{t^2}(b-a)|B|e^{-bt}  > 0
\eeq
We see from this equation that for $at \lesssim \ord (7)$ this extremum is a minimum (this is an approximate upper bound based on the assumption that $a \sim b$).  This should be compared with the upper bound obtained in our first model ($at < 11/2$).  We see that there is only marginal improvement our first model.  The gravitino mass is bounded from below by
\beq
m_{3/2} \sim \frac{|W_0|}{t^{3/2}} \gtrsim 5\times10^{-4}\; \text{M}_{\text{P}}\;\; \text{or}\;\; 5\times10^{14} \; \text{GeV} 
\eeq
Where, as before, $\xi\sim\ord(100)$. We may estimate the value of the scalar potential at the minimum by inserting the extremization equation (eqn.~\eqref{AW0_min}) into eqn.~\eqref{AKSRT_pot}.  This yields the following relation
\begin{eqnarray}\label{AKSRT_Vmin}
V|_{min} &=& -\left(4|W_0|\left(+\frac{27\xi |W_0|}{32t^{7/2}a} -\frac{b^2}{a}|B|e^{-bt}+b|B|e^{-bt}\right)\right)t^{-2} +\frac{3|W_0|^2\xi}{4t^{9/2}} \nonumber \\
         &=& \frac{3|W_0|^2\xi}{4t^{9/2}}\left(-\frac{9}{2at}+1\right) -\frac{4|W_0|}{t^2}|B|be^{-bt}\left(1-\frac{b}{a}\right)
\end{eqnarray} 
In principle, $V|_{min}$ can be fine-tuned to zero. Due to the transcendental nature of eqn.~\eqref{AKSRT_Vmin}, this has to be done numerically. We also note that, as with the first model, this model is, in principle, susceptible to destabilization via 1-loop quantum corrections (\`{a} la eqn.~\eqref{Vmin_Loop}).  However, with sufficiently large values of $s$, this correction can be suppressed and the classical minimum maintained.  

\bibliographystyle{hieeetr}	% (uses file "hieeetr.bst")
\bibliography{refs}		% expects file "refs.bib" 

\begin{thebibliography}{10}

\bibitem{Grana:2005jc}
M.~Grana, ``{Flux compactifications in string theory: A comprehensive
  review},'' {\em Phys. Rept.}, vol.~423, pp.~91--158, 2006, hep-th/0509003.

\bibitem{Douglas:2006es}
M.~R. Douglas and S.~Kachru, ``{Flux compactification},'' {\em Rev. Mod.
  Phys.}, vol.~79, pp.~733--796, 2007, hep-th/0610102.

\bibitem{Kachru:2003aw}
S.~Kachru, R.~Kallosh, A.~D. Linde, and S.~P. Trivedi, ``{De Sitter vacua in
  string theory},'' {\em Phys. Rev.}, vol.~D68, p.~046005, 2003,
  hep-th/0301240.

\bibitem{Brustein:2004xn}
R.~Brustein and S.~P. de~Alwis, ``{Moduli potentials in string
  compactifications with fluxes: Mapping the discretuum},'' {\em Phys. Rev.},
  vol.~D69, p.~126006, 2004, hep-th/0402088.

\bibitem{Bala:2005zx}
V.~Balasubramanian, P.~Berglund, J.~P. Conlon, and F.~Quevedo, ``{Systematics
  of Moduli Stabilisation in Calabi-Yau Flux Compactifications},'' {\em JHEP},
  vol.~03, p.~007, 2005, hep-th/0502058.

\bibitem{Conlon:2008wa}
J.~P. Conlon, A.~Maharana, and F.~Quevedo, ``{Towards Realistic String
  Vacua},'' {\em JHEP}, vol.~05, p.~109, 2009, 0810.5660.

\bibitem{Blumenhagen:2009gk}
R.~Blumenhagen, J.~P. Conlon, S.~Krippendorf, S.~Moster, and F.~Quevedo,
  ``{SUSY Breaking in Local String/F-Theory Models},'' {\em JHEP}, vol.~09,
  p.~007, 2009, 0906.3297.

\bibitem{deAlwis:2009fn}
S.~P. de~Alwis, ``{Classical and Quantum SUSY Breaking Effects in IIB Local
  Models},'' {\em JHEP}, vol.~03, p.~078, 2010, 0912.2950.

\bibitem{Baer:2010uy}
H.~Baer, S.~de~Alwis, K.~Givens, S.~Rajagopalan, and H.~Summy, ``{Gaugino
  Anomaly Mediated SUSY Breaking: phenomenology and prospects for the LHC},''
  {\em JHEP}, vol.~05, p.~069, 2010, 1002.4633.

\bibitem{Baer:2010rz}
H.~Baer, S.~Alwis, K.~Givens, S.~Rajagopalan, and W.~Sreethawong, ``{Testing
  the gaugino AMSB model at the Tevatron via slepton pair production},'' {\em
  JHEP}, vol.~01, p.~005, 2011, 1010.4357.

\bibitem{deAlwis:2005tf}
S.~P. de~Alwis, ``{Effective potentials for light moduli},'' {\em Phys. Lett.},
  vol.~B626, pp.~223--229, 2005, hep-th/0506266.

\bibitem{Gallego:2008qi}
D.~Gallego and M.~Serone, ``{An Effective Description of the Landscape - I},''
  {\em JHEP}, vol.~01, p.~056, 2009, 0812.0369.

\bibitem{Gallego:2009px}
D.~Gallego and M.~Serone, ``{An Effective Description of the Landscape - II},''
  {\em JHEP}, vol.~0906, p.~057, 2009, 0904.2537.

\bibitem{Gallego:2011jm}
D.~Gallego, ``{On the Effective Description of Large Volume
  Compactifications},'' {\em JHEP}, vol.~1106, p.~087, 2011, 1103.5469.

\bibitem{Brizi:2009nn}
L.~Brizi, M.~Gomez-Reino, and C.~A. Scrucca, ``{Globally and locally
  supersymmetric effective theories for light fields},'' {\em Nucl.Phys.},
  vol.~B820, pp.~193--212, 2009, 0904.0370.

\bibitem{Gray:2008zs}
J.~Gray, Y.-H. He, A.~Ilderton, and A.~Lukas, ``{STRINGVACUA: A Mathematica
  Package for Studying Vacuum Configurations in String Phenomenology},'' {\em
  Comput. Phys. Commun.}, vol.~180, pp.~107--119, 2009, 0801.1508.

\bibitem{Balasubramanian:2004uy}
V.~Balasubramanian and P.~Berglund, ``{Stringy corrections to Kahler
  potentials, SUSY breaking, and the cosmological constant problem},'' {\em
  JHEP}, vol.~11, p.~085, 2004, hep-th/0408054.

\bibitem{Westphal:2006tn}
A.~Westphal, ``{de Sitter String Vacua from Kahler Uplifting},'' {\em JHEP},
  vol.~03, p.~102, 2007, hep-th/0611332.

\bibitem{Becker:2002nn}
K.~Becker, M.~Becker, M.~Haack, and J.~Louis, ``{Supersymmetry breaking and
  alpha'-corrections to flux induced potentials},'' {\em JHEP}, vol.~06,
  p.~060, 2002, hep-th/0204254.

\bibitem{Rummel:2011cd}
M.~Rummel and A.~Westphal, ``{A sufficient condition for de Sitter vacua in
  type IIB string theory},'' 2011, 1107.2115.

\bibitem{Covi:2008ea}
L.~Covi {\em et~al.}, ``{de Sitter vacua in no-scale supergravities and
  Calabi-Yau string models},'' {\em JHEP}, vol.~06, p.~057, 2008, 0804.1073.

\bibitem{GomezReino:2006dk}
M.~Gomez-Reino and C.~A. Scrucca, ``{Locally stable non-supersymmetric
  Minkowski vacua in supergravity},'' {\em JHEP}, vol.~0605, p.~015, 2006,
  hep-th/0602246.

\bibitem{Berg:2005ja}
M.~Berg, M.~Haack, and B.~Kors, ``{String loop corrections to Kaehler
  potentials in orientifolds},'' {\em JHEP}, vol.~11, p.~030, 2005,
  hep-th/0508043.

\bibitem{Berg:2005yu}
M.~Berg, M.~Haack, and B.~Kors, ``{On volume stabilization by quantum
  corrections},'' {\em Phys. Rev. Lett.}, vol.~96, p.~021601, 2006,
  hep-th/0508171.

\bibitem{vonGersdorff:2005bf}
G.~von Gersdorff and A.~Hebecker, ``{Kaehler corrections for the volume modulus
  of flux compactifications},'' {\em Phys. Lett.}, vol.~B624, pp.~270--274,
  2005, hep-th/0507131.

\bibitem{Kaplunovsky:1993rd}
V.~S. Kaplunovsky and J.~Louis, ``{Model independent analysis of soft terms in
  effective supergravity and in string theory},'' {\em Phys. Lett.}, vol.~B306,
  pp.~269--275, 1993, hep-th/9303040.

\bibitem{Brignole:1997dp}
A.~Brignole, L.~E. Ibanez, and C.~Munoz, ``{Soft supersymmetry-breaking terms
  from supergravity and superstring models},'' 1997, hep-ph/9707209.

\bibitem{Grana:2003ek}
M.~Grana, T.~W. Grimm, H.~Jockers, and J.~Louis, ``{Soft Supersymmetry Breaking
  in Calabi-Yau Orientifolds with D-branes and Fluxes},'' {\em Nucl. Phys.},
  vol.~B690, pp.~21--61, 2004, hep-th/0312232.

\bibitem{Blumenhagen:2008aw}
R.~Blumenhagen, ``{Gauge Coupling Unification in F-Theory Grand Unified
  Theories},'' {\em Phys. Rev. Lett.}, vol.~102, p.~071601, 2009, 0812.0248.

\bibitem{deAlwis:2008kt}
S.~P. de~Alwis, ``{Mediation of Supersymmetry Breaking in a Class of String
  Theory Models},'' {\em JHEP}, vol.~03, p.~023, 2009, 0806.2672.

\bibitem{Burgess:2010sy}
C.~P. Burgess, A.~Maharana, and F.~Quevedo, ``{Uber-naturalness: unexpectedly
  light scalars from supersymmetric extra dimensions},'' {\em JHEP}, vol.~05,
  p.~010, 2011, 1005.1199.

\end{thebibliography}
\end{document}